\documentclass[%
 reprint,
 amsmath,amssymb,
 aps,
]{revtex4-2}

\usepackage{graphicx}
\usepackage{dcolumn}
\usepackage{bm}

\begin{document}

\preprint{APS/123-QED}

\title{Graph Theoretic Approach Identifies Critical Thresholds \\
at which Galaxy Filamentary Structures Form}

\author{Sophia-Gisela Strey}\email{Sophia.Strey@gmail.com}
\affiliation{
State University of New York at Stony Brook, Stony Brook, NY
 }%
 
\author{Alexander Castronovo}
\affiliation{
Florida Atlantic University, Boca Raton, FL
}

\author{Kailash Elumalai}
\affiliation{
Stanford University, Redwood City, CA 
}

\date{\today}

\begin{abstract}
Numerical simulations and observations show that galaxies are not uniformly distributed. In cosmology, the largest known structures in the universe are galaxy filaments formed from the hierarchical clustering of galaxies due to gravitational forces. These structures consist of “walls” and “bridges” that connect clusters. Here, we use graph theory to model the structures as Euclidean networks in three-dimensional space. Using percolation theory, cosmological graphs are reduced based on the valency of nodes to reveal the inner, most robust structural formation. By constraining the network, we then find thresholds for physical features, such as length-scale and density, at which galaxy filament clusters are identified. 
\end{abstract}
\keywords{Cosmology \and Graph Theory \and Percolation \and Galaxy Filaments}

\maketitle


\section{\label{sec:level1}Introduction}

Large-scale filamentary structures are thought to have evolved through gravitational instabilities and density fluctuations. There are two competing theories explaining the formation of galaxy clusters \cite{bond1996}: The Zeldovich theory describes primordial density fluctuation in the early universe causing the condensation of gas. Alternatively, the rival theory of hierarchical clustering suggests that merging halos explains the resulting Gaussian density fields and voids.

In 1984, researchers analyzed the Center for Astrophysics (CfA) I redshift catalog \cite{cfa1survey} using percolation analysis. Their findings indicated that these large-scale distributions were consistent with network structures \cite{einasto1984}. The next year, further work introduced the idea of using \textit{minimum spanning trees} (MST) \cite{barrow1985}, a concept borrowed from graph theory, providing a mathematical framework for representing relationships between objects.

These relationships are simplified into points (\textit{nodes}) and connections (\textit{edges}), weighted according to a numerical measure used to characterize their relationship. A set of nodes and edges makes up a graph. When a graph contains no circuits or closed loops, it is a tree. If a tree contains all of the nodes in a graph (rather than a subset), it is called a spanning tree since it spans the entire dataset\cite{essam1970}. A minimum spanning tree takes this one step further by minimizing total edge weights for the tree. Applied to cosmology, graph theoretic analyses\cite{barrow1985} identify the dominant pattern of connectedness in a set of points where each represents a galaxy. The resulting skeletal pattern can be analyzed in a quantitative manner, which differs from other criteria previously used to characterize the geometry of galaxy clustering. This quantitative measure, in turn, serves as an objective way to identify filaments. Given the statistical likelihood that all edges have different weights, there will be only one unique MST in any point dataset since there will only be one shortest-path solution \cite{barrow1985}.

While previous work has been valuable in advancing our understanding of the filamentary structure between clusters, the relationship between the characteristic thread-like geometry of filaments and the conditions of the gravitational collapse of matter into those filaments is not yet known. To investigate this question, we will be looking for the critical distance at which the galaxy clusters begin to form a filamentary structure over time. 

\section{\label{sec:level1}Approach}

If we define the filamentary structure of a galaxy cluster using its minimum spanning tree, we can say that it remains \textit{functional} even after undergoing some operation, as long as the tree is preserved and retains its tree-like properties. The failure of key points and bonds in a network will cause irreparable losses and changes in the structure of the system. By identifying these key points, we can not only evaluate the stability of the system but also determine which substructures are most critical, i.e., galaxies most vital to the preserved functionality of the filament. In other words, the most critical points are galaxies where, without their existence, the filamentary structure will break down.

To identify filament \textit{dynamics}, we have to look at time. We can do this by looking at different radial distances from Earth, showing us different points in cosmological time. Looking at how the robustness of galaxy filaments evolves and changes under different constraints, we can try to find under what circumstances galaxy filaments begin to form. 

\begin{figure*}
\resizebox{.7\linewidth}{!}{\includegraphics{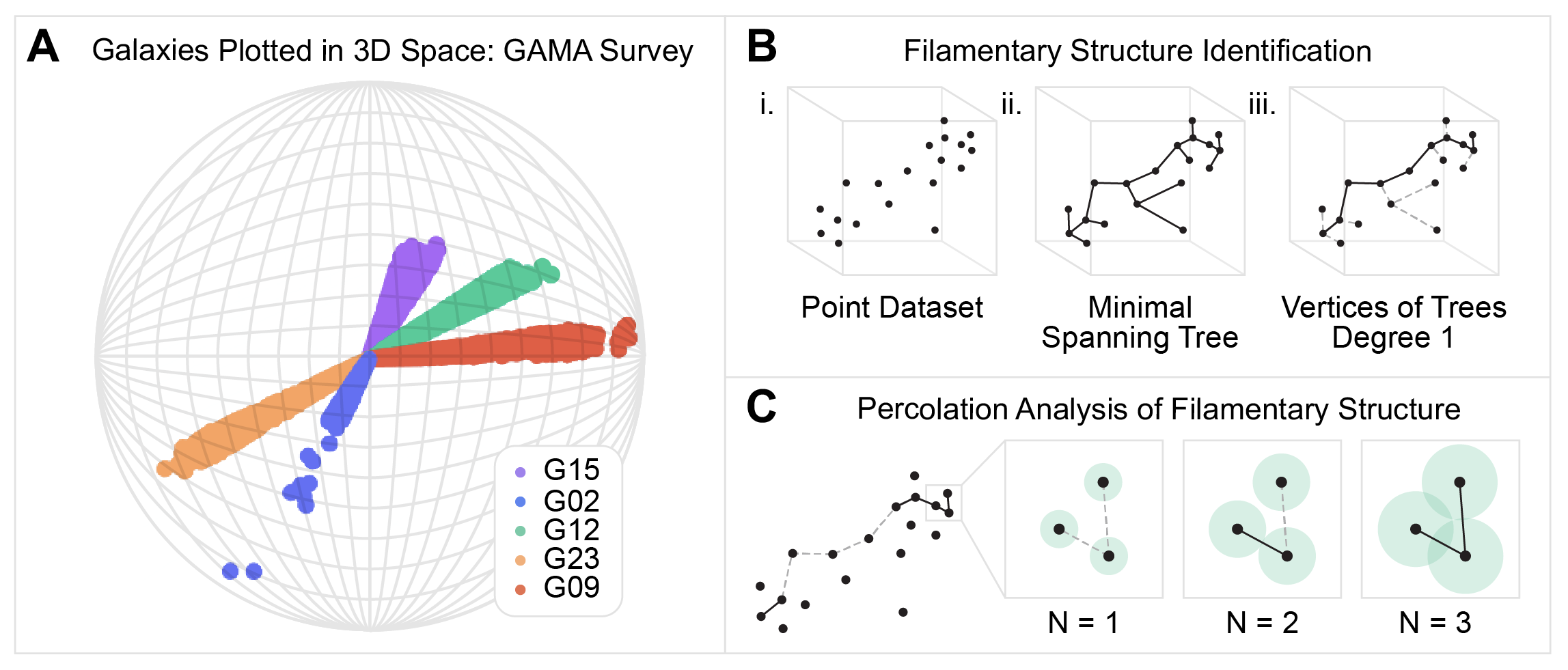}}
\caption{\label{fig:wide} Approach to identification and analysis of galaxy filaments. (A) GAMA Dataset Sections, G15, G02, G12, G23, G09, described in Table I represented in 3-dimensional space. (B) The identification of filamentary structure by constructing and reducing a minimum spanning tree the point dataset described in A. (C) The percolation analysis performed on the identified filamentary structure by iterating the critical length scale to form a bond. }
\label{fig:approach}
\end{figure*}

Recently, with the Data Release 4 of the Galaxy and Mass Assembly (GAMA) spectroscopic survey in 2022, there have been many cosmological analyses done due to the inclusion of redshift estimation measurements on galaxy candidates \cite{driver2022gama}. Cosmological redshift is a useful measure because it describes the Doppler shift light as it travels through space. Using Hubble’s Law and the Cosmology.jl Julia package, we can find the radial distance of galaxies. Therefore, a critical next step is to characterize this structure (Figure \ref{fig:approach}).

Relating to cosmology and galaxy filaments, graphs can tell us the underlying structure of the filaments and their associated features (walls, bridges, clusters, and voids) \cite{barrow1985}. The robustness of a graph can be determined when a point data set can no longer be represented by the original minimum spanning tree \cite{yang2021}. The current study proposes to assess the robustness of the galaxy cluster by finding the \textit{percolation threshold}, a measure adapted from statistical physics. The aim of percolation analysis is to study the emergence of clusters, or any ordered structure, from a disordered one. This is done by varying the probabilities of occupancy when a transition occurs. As the probabilities increase, the size of the largest cluster grows until it spans the entire system: a fully percolated system \cite{Lee2016}. Using distance as a proxy for occupational probability, we define the probability of connecting two sites by the power-law as a function of their distance ($y=ax^k$ where $a$ controls the strength of distance dependence).

We implemented all algorithms and analysis in the Julia programming language. Julia is a just-in-time compiled language optimized for high-performance computing.  As such, it has emerged as one of the best choices for dynamic and graph theoretical modeling of data because of its speed (as fast as C) as well as its support from the community in developing state-of-the-art computational tools\cite{bezanson2017}. Specifically, we used the following Julia packages for our analysis: SimpleWeightedGraphs.jl and Cosmology.jl.

In this study, we assume scale invariance in galaxy clustering \cite{LABINI199861}. Analysis of the CfA, Perseus-Pisces, SSRS, IRAS, etc., shows that galaxy structures are ``highly irregular and self-similar". However, due to a lack of available data, the evidence of scale invariance becomes weaker at a scale of greater than 150 Mpc.

\section{\label{sec:level1}Data}
The GAMA Data Release 4, published in March 2022, provides over 300,000 galaxy spectroscopic redshift samples over $250\text{deg}^2$ of sky. These samples were taken across five sky regions with bounds as shown in Table \ref{table:sky-regions}. The G02 sky region is not used in this study because it is incomplete \cite{driver2022gama}.

\begin{table}[!h]
    \begin{center}
        \begin{tabular}{c|c|c}
            Region Label & RA Range & Dec Range\\
            G02 & 30.2 to 38.8 & -10.25 to 3.72\\
            G09 & 129.0 to 141.0 & -2.0 to 3.0\\
            G12 & 174.0 to 186.0 & -3.0 to 2.0\\
            G15 & 211.5 to 223.5 & -2.0 to 3.0\\
            G23 & 338.1 to 351.9 & -35.0 to -30.0\\
        \end{tabular}
        \caption{GAMA regions and their right ascension and declination bounds.}

        \label{table:sky-regions}
    \end{center}
\end{table}

This survey includes the galaxy's stellar mass function and its sub-division by morphological type. This function is defined as the number density of galaxies in a selected mass interval. Redshift measurements have exceptionally high completeness ($\leq 95\%$) and include many low redshift populations ($z\leq 0.25$), as seen in Figure 1A \cite{driver2022gama}. Completeness refers to whether the dataset contains all galaxies within the relevant regions of space, considering sensitivity, resolution, and error. Figure 1A shows a scatter plot showing the distribution of galaxies in each region. The \textit{x}, \textit{y}, and \textit{z} coordinates were calculated as illustrated in Figure 1A and are in the units of Mpc. 

\section{\label{sec:level1}Methods}

Our study looks at various subsections of the data described in the previous section. The methods described were applied in the same way to each subsection of the data. The Methods Section follows the order of the Julia scripts that can be found in our publicly available repository. 

\subsection*{Initial Data Processing} We treat the GAMA dataset as a three-dimensional point dataset using right ascension and declination measures. This initial processing converts spherical coordinates into Cartesian coordinates in units of Mpc. Using the Cosmology.jl package, the comoving radial distance is calculated for each galaxy based on the Dimensionless Hubble constant ($H_0 = 0.7$), matter density ($\Omega_M = 0.3$), and radiation density ($\Omega_R=0$) parameters of the cosmology model along with their cosmological age. A custom struct, ‘GalaxyDistance’, is defined to store the distance information between pairs of galaxies and their indices. This also filters out null values. 

\subsection*{Filamentary Structure Identification} Using ‘GalaxyDistance’ as its input, we create a weighted graph. From this graph, we construct an MST using Krustal’s Algorithm. For a weighted graph $G = (V, E, w)$, where $V$ is the set of nodes, $E$ is the set of edges, and $w$ is the weight for each edge connecting node $u$ to node $v$, the MST is defined as the subset of edges, $T$, whose weight 
$$w(T) = \sum_{(u,v)\in T} w(u,v)$$
is minimized. Krustal’s Algorithm sorts edges in the graph in increasing order of their weight and iterates through the sorted edges. For each edge, it is checked if adding it to the current MST will create a cycle. The process continues until $V-1$ edges are included in the MST, where $V$ is the number of vertices. This algorithm has a time complexity of $O(\log E)$. Finally, all nodes of degree one connected to nodes with degrees exceeding two are removed along with their offshooting branches. This is how we define the \textit{functional} filamentary structure. 

The same struct is redefined using only edges from the MST, and edges are sorted by distance.  Note that for the purposes of this study, we will assume all edges have different lengths because edge weights are calculated with 16 digits of precision (limited by limitations of a 64-bit floating point value), so it is highly unlikely two edges will have the same weight. However, if two edges do have the same weight, then it is most likely not going to significantly affect the results of the analysis since the trees will be similar in structure.

\subsection*{Percolation} We implement the Newman-Ziff Monte Carlo Algorithm for bond percolation. Unlike traditional percolation algorithms where cluster growth is typically simulated, this algorithm utilizes the fact that the probability of a bond being part of the largest cluster in a percolating system is a continuous function of the site occupation probability. Simulations can be computationally intensive, especially for a large system.  For this reason, the Newman-Ziff method is preferable. The method works by starting with an empty graph and sequentially adding bonds with a given occupation probability until all points are added to the cluster. In a cosmological system, distance can be used as a proxy for occupation probability. In this case, the Newman-Ziff method can be applied by constructing a curve of the probability that the galaxy belongs to the largest connected component as a function of distance. The threshold or critical point at which percolation occurs can be determined by finding the point on this curve where the probability of belonging to the largest cluster/component changes from zero to some nonzero value \cite{newman2001}. 

The percolation structure that we define represents a model of percolation in a grid. The grid contains ‘true’ (open) and ‘false’ (closed) cells. The ‘QuickUnion’ structures (‘wuf1’ and ‘wuf2’) are used to keep track of connectivity in the grid. The results have a $95\%$ confidence interval. 

\section{\label{sec:level1}Results}

\begin{figure*}
\begin{center}
\resizebox{.5\linewidth}{!}{
\includegraphics{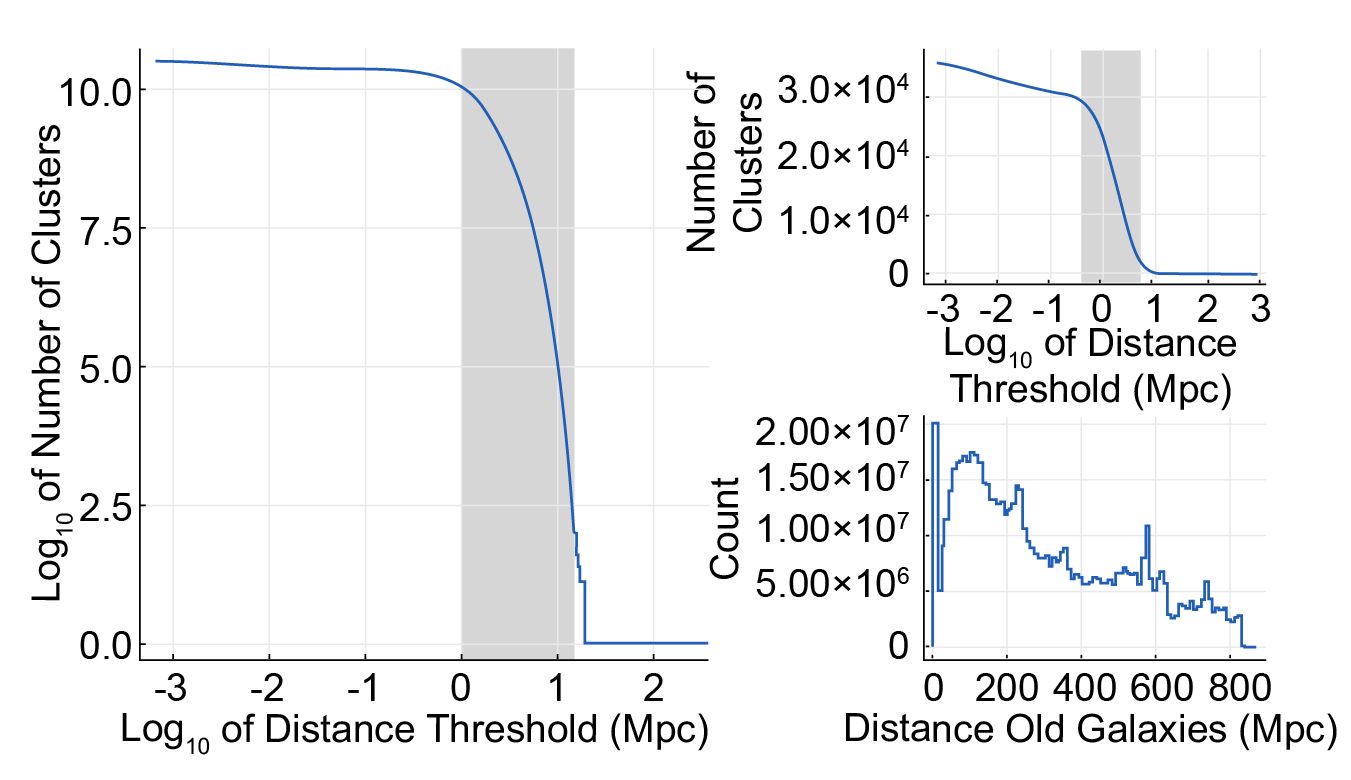}
}
\caption{\label{fig:wide} Percolation analysis on older galaxies with logarithmic scaling (G15). Length scale percolation threshold = ~0.7 Mpc.}
\end{center}
\end{figure*}

\begin{figure*}
\begin{center}
\resizebox{.5\linewidth}{!}{
\includegraphics{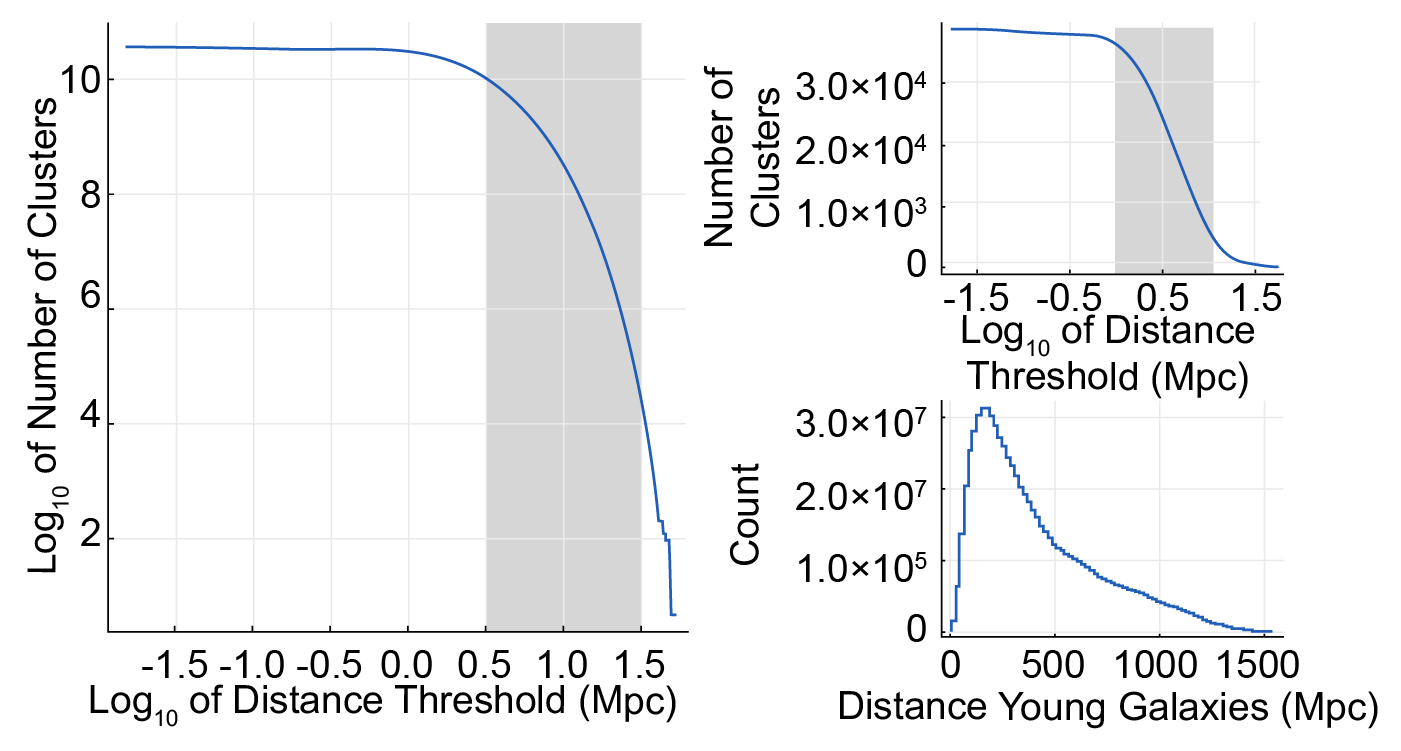}
}
\caption{\label{fig:wide} Percolation analysis on younger galaxies with logarithmic scaling (G15). Length scale percolation threshold =  ~1.2 Mpc. }
\end{center}
\end{figure*}

\begin{figure*}
\begin{center}
\resizebox{.5\linewidth}{!}{
\includegraphics{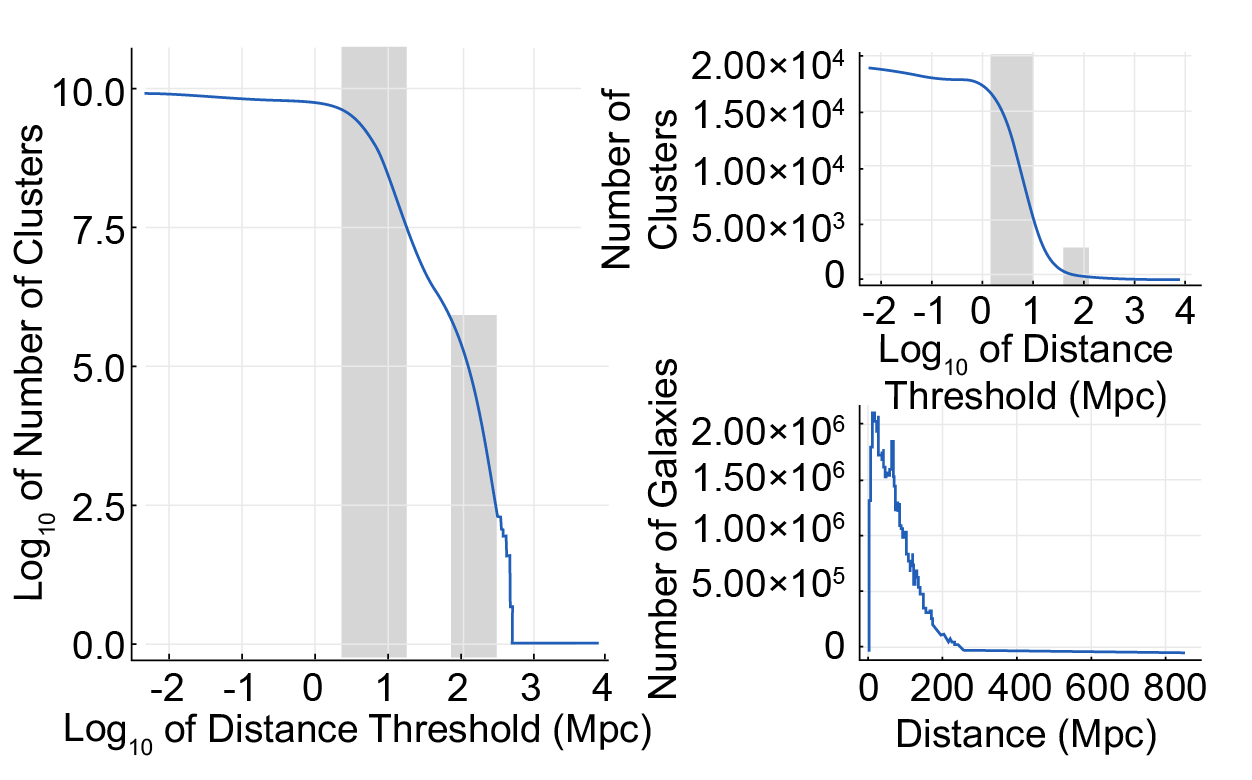}
}
\caption{\label{fig:wide} Percolation analysis on all galaxies (¼ of total dataset given computation constraints).}
\end{center}
\end{figure*}

\begin{figure*}
\begin{center}
\resizebox{.5\linewidth}{!}{
\includegraphics{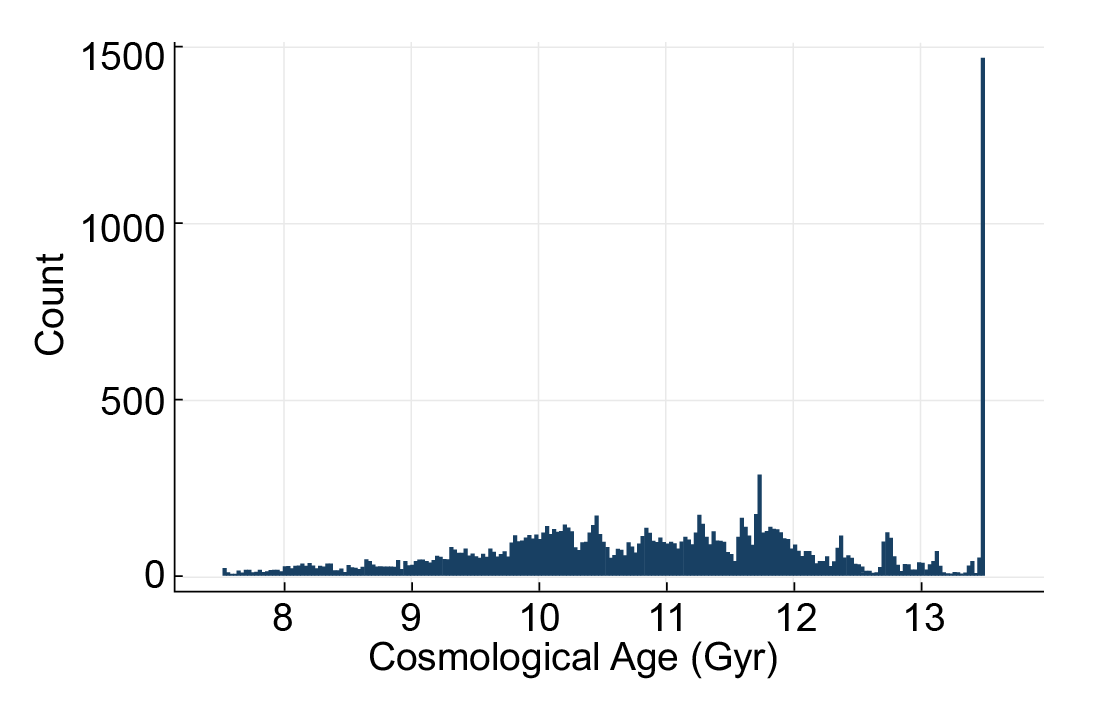}
}
\caption{\label{fig:wide} Histogram of Cosmological times of whole datasets in Gyr.}
\end{center}
\end{figure*}

\subsection*{Younger galaxies have a larger critical length-scale.} The critical length-scale measures of older galaxies ($\geq 11$ Gyr) = $\approx 0.7$ Mpc (Figure 2) and younger galaxies ($\leq 11$ Gyr) = $\approx 1.2$ Mpc (Figure 3). The threshold of older galaxies lies between 0.2 and 1.3 Mpc, and the threshold of younger galaxies lies between 0.5 and 1.7 Mpc. Thus, older galaxies have a sharper transition than younger galaxies. Note that the figure has been logarithmically scaled. 

\subsection*{Two critical points in Total Sections.} As shown in Figure 4, the percolation threshold of the total section of G15 occurs at two points: ~0.8 Mpc and ~2.5 Mpc. Having two percolation thresholds means that there are two distinct scales at which clustering properties emerge in the galaxy dataset. The first has a sharper transition than the second. This can be seen across all remaining sections of the GAMA dataset (excluding G02 for being incomplete).

Additionally, there is little to no difference between the percolation thresholds or curves of younger and older galaxies when percolating over only the constructed galaxy filament. Once again, this can be seen across all remaining sections of the GAMA dataset (excluding G02 for being incomplete). Figure 2, showing results for G15, is representative. Percolation curves for other sections (G12, G23, G09) can be found in our code repository. 

\section{\label{sec:level1}Discussion}

\subsection*{Dimensions of a Filament} Our results, showing two distinct percolation thresholds, can be interpreted as two length-scale transitions of a shape defined by two-dimensional measurements. Generalizing filament shapes to a cylinder, these transitions could represent the diameter and length of its geometry. However, the observation of two distinct percolation thresholds prompts an alternative interpretation related to the age diversity of these thresholds. Rather than attributing these thresholds to a continuous filament with specific dimensions, we propose that they correspond to the evolutionary stages of galaxy filaments based on age. 

The sharper transition associated with the first threshold suggests the presence of dominant clustering structures at a smaller scale, likely indicative of characteristics exhibited by younger galaxies. Conversely, the second threshold, marked by a less abrupt transition, might be indicative of a larger-scale clustering phenomenon, potentially corresponding to older galaxies. This is supported by the interval of each transition and its correspondence to the distance thresholds found in the entire dataset. These transitions, signaling shifts between disconnected and connected states, unfold in distinct stages, mirroring the evolution of galaxies over time.

\subsection*{Difference Between Entire Dataset versus Younger and Older Subsections} 
Our results seem to imply that different age groups of galaxies undergo different evolutionary processes. The GAMA dataset contains a mix of galaxies with varying properties. The existence of two thresholds solely within the entire dataset suggests that both older and younger galaxies exhibit a higher degree of homogeneity within their respective groups. This challenges the notion of a continuous filament with specific dimensions and instead supports the idea that these thresholds delineate transitions between age-dependent characteristics. The distinct thresholds may reflect the intricate interplay of various factors, such as matter distribution, density, and structural complexities, which evolve differently in galaxies of varying ages.

\section*{Code}  All code utilized in this manuscript may be found at:  https://github.com/tstrey/oastr2-leftovers.

\section*{Acknowledgments}
GAMA is a joint European-Australasian project based around a spectroscopic campaign using the Anglo-Australian Telescope. The GAMA input catalog is based on data taken from the Sloan Digital Sky Survey and the UKIRT Infrared Deep Sky Survey. Complementary imaging of the GAMA regions is being obtained by a number of independent survey programs, including GALEX MIS, VST KiDS, VISTA VIKING, WISE, Herschel-ATLAS, GMRT, and ASKAP, providing UV to radio coverage. GAMA is funded by the STFC (UK), the ARC (Australia), the AAO, and the participating institutions. The GAMA website is \href{http://www.gama-survey.org}{http://www.gama-survey.org}. We thank Hillel Sanhedrai, Rostam Razban, and Kalee Tock for providing feedback on the manuscript and Annabel Driussi for her assistance with producing figures. 

\break

\bibliography{references}
\end{document}